  \documentclass[pdf,aps,pra,twocolumn,notitlepage,floatfix, superscriptaddress, nofootinbib, letterpaper,longbibliography]{revtex4}
 
  \usepackage[pdftex, breaklinks, plainpages=false,colorlinks=true,citecolor=blue,linkcolor=blue,urlcolor=blue,filecolor=green,bookmarksopen=true]{hyperref}
  \usepackage{amsmath, amstext, amssymb, amsfonts,amsmath, mathtools}
  \usepackage{natbib, url, textcase, bm} 
  \usepackage{microtype, graphicx, algorithm, algpseudocode}
  \usepackage{subfigure}
  \usepackage[english]{babel}
  \usepackage{exscale}
  \usepackage[dvipsnames]{xcolor}
 \usepackage{setspace}          
 \usepackage[thicklines]{cancel}
  \graphicspath{ {./FIG/} }

 \newcommand{\CBPF}{Centro Brasileiro de Pesquisas F\'isicas, Rua Dr.\ Xavier Sigaud 150, 22290-180, Rio de Janeiro, Brazil}

\begin{document}

\title{ Process Tomography of Robust Dynamical Decoupling with Superconducting Qubits}

 \author{Alexandre M. Souza}  \email{amsouza@cbpf.br}   \affiliation{\CBPF}

\begin{abstract}
Dynamical decoupling is a technique that protects qubits against noise. The ability to preserve quantum coherence in the presence of noise is essential for developing quantum devices. Here, the Rigetti quantum computing platform was used to test different dynamical decoupling sequences {\bf in a single qubit}. The performance of the sequences was characterized by quantum process tomography and analyzed using the quantum channels formalism.  Pulse imperfections are shown here to limit the performance of dynamical decoupling on the Rigetti's qubits. However, the performance can be improved using robust dynamical decoupling, i.e., sequences that are robust against experimental imperfections. The sequences tested here outperformed previous dynamical decoupling sequences tested in the same platform.

\end{abstract}
\maketitle

\section{Introduction} \label{intro}

The past decade has seen great effort in developing quantum technologies, such as quantum computers, sensors, and memories. Notable devices are the quantum computers remotely accessible to the public via cloud services. As with any quantum system,
these computers are subject to errors arising from unavoidable interactions with the environment or control imperfections. The currently available quantum computers are subject to a high level of noise. Therefore, these computers are non-fault-tolerant machines, usually referred to as noisy intermediate-scale quantum computers~\cite{preskill}. To make the computation reliable and efficient, quantum error correction (QEC) codes must be used \cite{campbell}. The theory of QEC states that fault-tolerant quantum computations of arbitrary length can be achieved provided
that the error per gate remains below some threshold \cite{zurek} and high-fidelity auxiliary states are prepared \cite{kitaev,souza0}. However, QEC requires many auxiliary qubits, and practical implementation of QEC codes in current quantum computers remains a challenge. Therefore, methods for reducing the noise level and control imperfections are desirable.

Dynamical decoupling (DD) is a widely used technique for avoiding decoherence in quantum systems. The decoupling approach was originally developed in the framework of nuclear magnetic resonance
(NMR) \cite{DD,suterRMP}. In standard DD, a sequence of $\pi$-rotation pulses is periodically applied to a quantum system to attenuate the system-environment interaction. Experimental tests of DD implemented in different types of qubits have demonstrated the increase in the coherence times by several orders of magnitude \cite{biercuk,du,delange,ryan,sagi,saeedi,foton}. In nuclear spins in rare-earth-doped crystals, for example, a quantum memory with six-hour coherence time was achieved using decoupling \cite{zhong}. Apart from preserving the state of a quantum memory, DD can also be used to implement decoherence protected quantum gates \cite{sar,rong,barthel,souza1,souza2,souza3}, to enhance the sensitivity of quantum sensors \cite{taylor,cooper} and to characterize the noise spectrum \cite{noise1,noise2,noise3}. Earlier experiments also showed that the main limitation to DD is pulse imperfections \cite{souza4}.

Using the IBM and Rigetti platforms, \cite{lidar} demonstrated that DD can protect quantum states of individual qubits and entangled two-qubit states. This work used the simplest universal decoupling sequence, namely, the $XY-4$. This sequence can cancel perturbations of the most general environment up to the first order \cite{DD}, but it has little robustness against experimental imperfections \cite{souza4}. 

Here, we use the Rigetti computing platform to test different robust DD sequences in a single superconducting qubit. We show that the application of $XY-4$ can extend the lifetime of a qubit, but the $XY-4$ performance is limited because of pulse imperfections, whereas robust DD usage can correct the pulse errors.  We also have shown that the action of DD on the qubit dynamics cannot be understood as a simple modification of the qubit coherence time.

This paper is organized as follows. In Sec. \ref{secdd}, robust dynamical decoupling sequences are introduced. In Sec. \ref{methods}, the methods used are presented, and in Sec. \ref{results}, the results obtained are shown. In the last section, we draw conclusions.

\section{Robust Dynamical Decoupling }  \label{secdd}
When a qubit is subjected to a static magnetic noise, we can preserve the qubit state by refocusing its evolution using a Hahn echo \cite{hahn}, i.e., placing a single rotation by $\pi$ in the middle of the evolution. If the magnetic field fluctuates in time, the Hahn echo must be replaced by DD. The simplest sequence that can refocus magnetic fields fluctuating along the three coordinate axes is the $XY-4$, defined as the repeated application of the basic block $[ \tau/2 -X -\tau- Y- \tau- X -\tau- Y -\tau/2 ]$, where $X$ and $Y$ are rotations by $\pi$ around the $x$ and $y$ axes, and $\tau$ represents a time when the qubit evolves freely. Here, the coordinate axes are defined as the axes of the qubit Bloch sphere.

The $XY-4$ sequence has limited decoupling performance and robustness against experimental imperfections. However, its performance can be improved by different strategies, such as concatenated dynamical decoupling \cite{cdd,cdd2,cdd3}, in which the basic $XY-4$ block is recursively concatenated. Another strategy was introduced in early NMR experiments \cite{maudsley,conradi} and consists of combining one basic building block with a symmetry-related copy. The sequences built within this approach, such as the $XY-8$ and $XY-16$, exhibit good robustness against the most common experimental imperfections \cite{souza4}. 

Robust DD sequences that are not related to the basic $XY-4$, such as the Knill dynamical decoupling (KDD) \cite{souza5} and universally robust (UR) sequences \cite{genov}, can also be built. The KDD sequence is built from composite pulses, which are sequences
of consecutive pulses designed to cancel the most common systematic errors. Therefore, the robustness of the KDD sequence is not achieved because of a form of concatenation. The KDD sequence has been demonstrated to be extremely robust against flip-angle errors and off-resonance errors \cite{souza5}. The UR sequences are built from a Taylor expansion of the infidelity, which quantifies how far the DD propagator is from a target propagator. The expansion is taken with respect to the imperfect control parameter, whereas the pulse phases are chosen to nullify the series coefficients up to the largest possible order. Considering a DD cycle with $n$ pulses, where $n$ is even, a DD sequence robust to flip-angle errors can be found whose $kth$ pulse phase is given by
\begin{eqnarray} 
  \phi_k^{(n)} = (k-1)(k-2)\frac{\Phi^{(n)}}{2}+(k-1)\phi_2,
 \end{eqnarray}
    where 
    \begin{eqnarray} 
  \Phi^{(4m)} = \pm \frac{\pi}{m},  \Phi^{(4m+2)} = \pm \frac{2m\pi}{2m+1} , 
  \end{eqnarray}
 $k=1, \cdots, n$, and $\phi_2$ is an arbitrary phase. When we choose, for example, $n = 4$ and $\phi_2 = \pi/2$, we have $m =1$; in this case, the $XY-4$ sequence is recovered. High order sequences can be obtained by increasing the number of pulses.
   
\begin{figure}
\centering
\includegraphics[width=1 \linewidth]{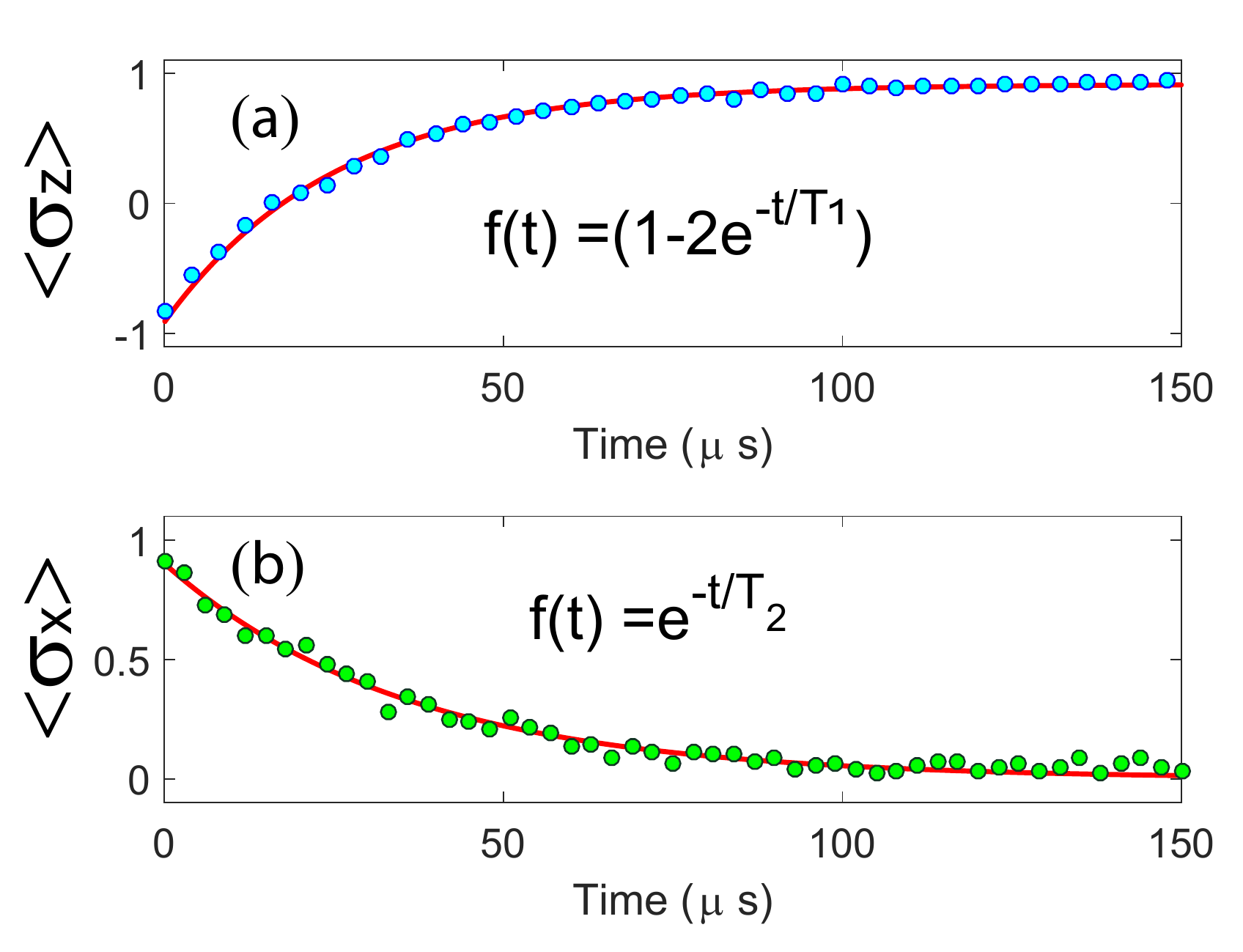}
\caption{Typical relaxation curves obtained in an inversion recovery experiment (a) and a Hahn echo experiment (b). The experiments were performed in the Rigetti quantum platform. The relaxation times were determined as $T_1 =25 \mu s$ and $T_2 = 35 \mu s$.        
\label{t1t2}}
\end{figure}

\section{ Methods}  \label{methods}

The performance of the sequences must be quantified regardless of the initial state, which is usually not known in most quantum information applications. A good quantifier for this case is the process fidelity \cite{wang}
\begin{equation}
\mathcal{F} =\frac{|\chi_{t} \chi_{dd} ^\dagger|}{\sqrt{Tr(\chi_{t} \chi_{t}^\dagger) Tr(\chi_{dd} \chi_{dd}^\dagger)} }. \label{fidel}
\end{equation}

Here, $\chi_{dd}$ is the process matrix corresponding to the evolution of the qubit when DD is applied, and $\chi_{t}$ is the process matrix corresponding to a target process. The $\chi$-matrix can be used to characterize non-unitary dynamics $\rho_f = \sum_{nm} \chi_{nm} E_n \rho_i E_m^{\dagger}$, where $\rho_i$ and $\rho_f$ are the density matrices at the beginning and end of the evolution, respectively. The set of operators 
$\{E_m\}$ form a basis. Here, we choose the basis set as $\{ I, \sigma_x, i\sigma_y\, \sigma_z\}$, where $I$ is the $2$ $\times$ $2$ identity matrix and $\sigma_k$ is one of the Pauli matrices. Using this basis, the process matrix $\chi_{dd}$ can be determined via quantum process tomography \cite{nielsen}, whereas the target matrix, $\chi_{dd}$, is given by

\begin{eqnarray}
\chi_{t} &=& \begin{bmatrix}1 & 0 &  0 & 0 \\ 0 & 0 &  0 & 0 \\ 0 & 0 &  0 & 0 \\ 0 & 0 &  0 & 0 \end{bmatrix}. \label{ident}
\end{eqnarray}

To analyze the qubit dynamics, we can use the quantum channels formalism. The density matrix
evolution for a given quantum channel is given by
\begin{equation}
\rho_f = \sum_n K_n \rho_i K_n^{\dagger}, \label{krauss}
\end{equation}
where $K_n$ is a Krauss operator \cite{nielsen}. To describe decoherence and relaxation processes of a qubit, the amplitude damping (AD) channel, which describes the dissipative interactions between the system and its environment, and the phase damping (PD) channel, which models the loss of coherence without loss of energy, are often used. The Krauss operators corresponding to the simultaneous action of AD and PD can be written as  
\begin{eqnarray} 
K_1  =  \sqrt{\lambda} \begin{bmatrix} 1 &  0  \\ 0 & \sqrt{1-\gamma}\end{bmatrix}, K_2 = \sqrt{1-\lambda} \begin{bmatrix} 0 &  \sqrt{\gamma}  \\ 0 & 0  \end{bmatrix}  \nonumber \\
K_3  =  \sqrt{1-\lambda}   \begin{bmatrix} 1 &  0  \\ 0 & \sqrt{1-\gamma}\end{bmatrix}, \text{ and } K_4  =  \sqrt{\lambda} \begin{bmatrix} 0 &  \sqrt{\gamma}  \\ 0 & 0  \end{bmatrix},  
\label{PDGAD} \end{eqnarray} 
where $\gamma = (1- e^{\Delta t \beta})$ is the probability that the excited state has decayed to the ground state in the time interval $\Delta t$, and $1-\lambda = (1-e^{-\Delta t \alpha})/2$ is the probability that the phase $\phi$, in the general state $|\psi\rangle =a |0\rangle +e^{ i \phi} b |1\rangle$, has inverted by interacting with the environment. According to this model, the coherence decays exponentially with decay constant $T_2 = (\alpha + \beta/2)^{-1}$, whereas all initial states are mapped to the ground state $|0\rangle$. For numerical calculations, the evolution of the qubit was discretized into small time steps, $\Delta t$, and the equations \eqref{krauss} and \eqref{PDGAD} were applied recursively. 

Usually, when applying DD sequences, we aim to remove dephasing. If all dephasing processes are removed ($\alpha =0$), then we have $T_2 = 2T_1$, where $T_1 =1/\beta$ is the longitudinal relaxation time. Figure \ref{t1t2} shows the results of two experiments performed to measure the relaxation times of a qubit. The transverse relaxation time was measured by a spin echo \cite{hahn,nmrbook} sequence as $T_2 = 35 \mu s$, and the longitudinal relaxation time was measured by an inversion recovery sequence \cite{nmrbook} as $T_1 = 25 \mu s$.

\begin{figure*}[t!]
\centering
\includegraphics[width=1 \linewidth]{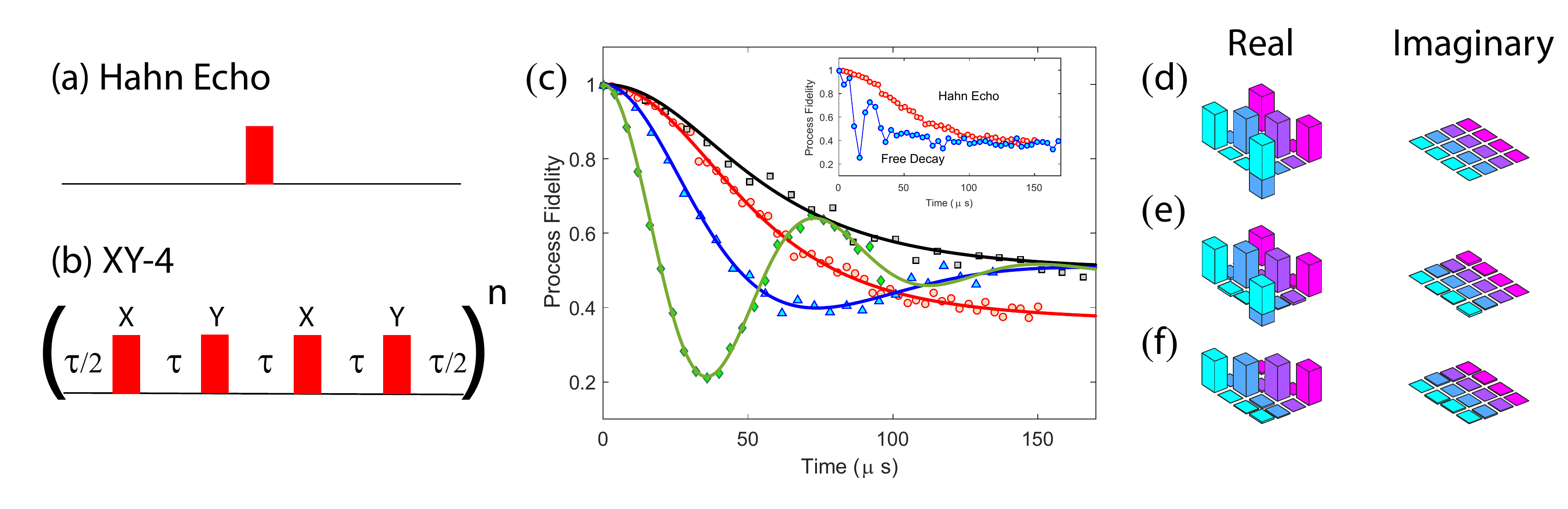}
\caption{ Comparisons between the Hahn echo, the $XY-4$ sequence, and the free evolution. (a) Illustration of the Hahn echo using a single $\pi$ pulse represented by the red rectangle placed in the middle of the evolution. (b) Illustration of the $XY-4$ sequence; the free evolution is replaced by $n$ applications of the basic $XY-4$ cycle. (c) Process fidelity as a function of time for the Hahn echo (red circles) and the $XY-4$ sequence. The time intervals between the $XY-4$ pulses are $\tau =120 ns$ (black squares), $\tau = 80 ns$ (blue triangles), and $\tau = 40 ns$ (green diamonds). The solid lines are the fitted models (see text). The inset compares the Hahn echo and the free evolution of the qubit. (d) The theoretical process matrix representing the quantum channel described by the equations \ref{PDGAD} when the evolution time $t \rightarrow \infty$. In (e) and (f), we can see the process matrices observed at $t = 150 \mu s$ for the Hahn echo and the $XY-4$ sequence, respectively.
\label{echo}}
\end{figure*}

\section{Results} \label{results} Using the above formalism, quantum process tomographies of different DD sequences were performed in the Rigetti platform. For each DD tested, many DD cycles were applied to a single qubit in the Aspen-4-2Q-H lattice. After a defined number of cycles, a quantum process tomography was performed and the process fidelity was computed to quantify the performance of the sequence.

\begin{figure}
\centering
\includegraphics[width=1.0 \linewidth]{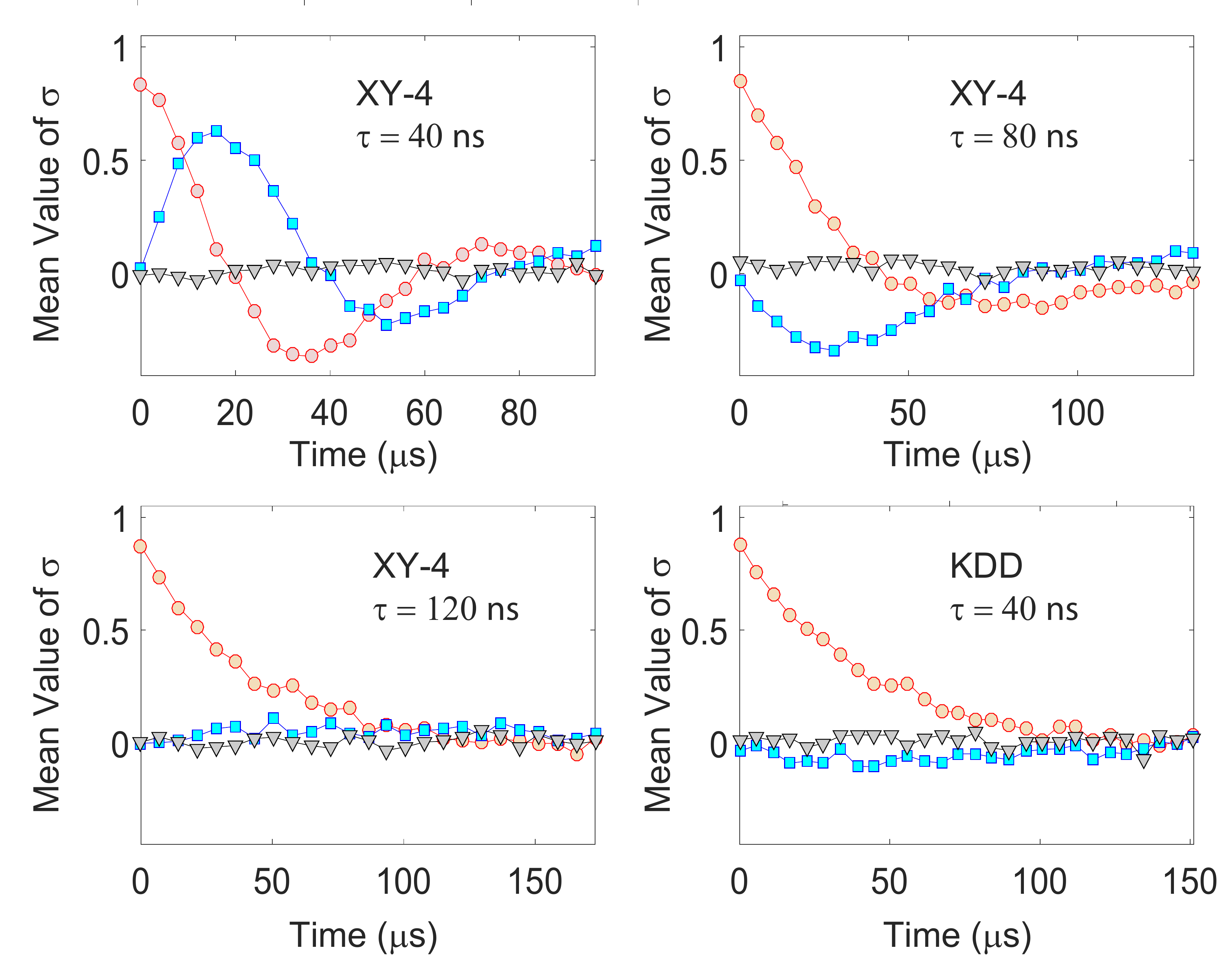}
\caption{Mean value of the operators $\sigma_x$ (red circles), $\sigma_y$ (blue rectangles), and $\sigma_z$ (black triangles) as a function of time. The initial state is $(|0\rangle + |1\rangle)/\sqrt{2}$.  
\label{oscilation}}
\end{figure}

First, the performance of the basic $XY-4$ was studied. Figure \ref{echo} compares the $XY-4 $ sequence, the Hahn echo, and the free evolution, i.e., when no DD is applied to the qubit. Under free evolution, the fidelity oscillates at the beginning and decays to a constant value. These oscillations are corrected when the Hahn echo is used, which indicates that a spurious and small DC magnetic field might be present. The fidelity is improved by applying the $XY-4$ sequence when the delay between pulses is $\tau = 120 ns$. However, when the delay is decreased to $\tau = 80 ns $ and $\tau = 40 ns$, the fidelity decreases quickly to a minimum value and then increases until finally stabilizing at a constant value. Ideally, decreasing the delay between pulses should improve decoupling. However, when errors are present, smaller delays also imply that the error will accumulate faster, resulting in an unwanted effect. The recurrence in the fidelity decay observed for the $XY-4$ sequence is similar to results obtained in NMR experiments \cite{souza6}, which shows that, in the presence of pulse errors, the $XY-4$ sequence can generate an additional effective magnetic field that rotates the qubit. Figure \ref{oscilation} shows that the angular frequency, $\omega$, increases when $\tau$ decreases. When robust sequences are used, the oscillation is not observed.

\begin{figure}
\centering
\includegraphics[width=1 \linewidth]{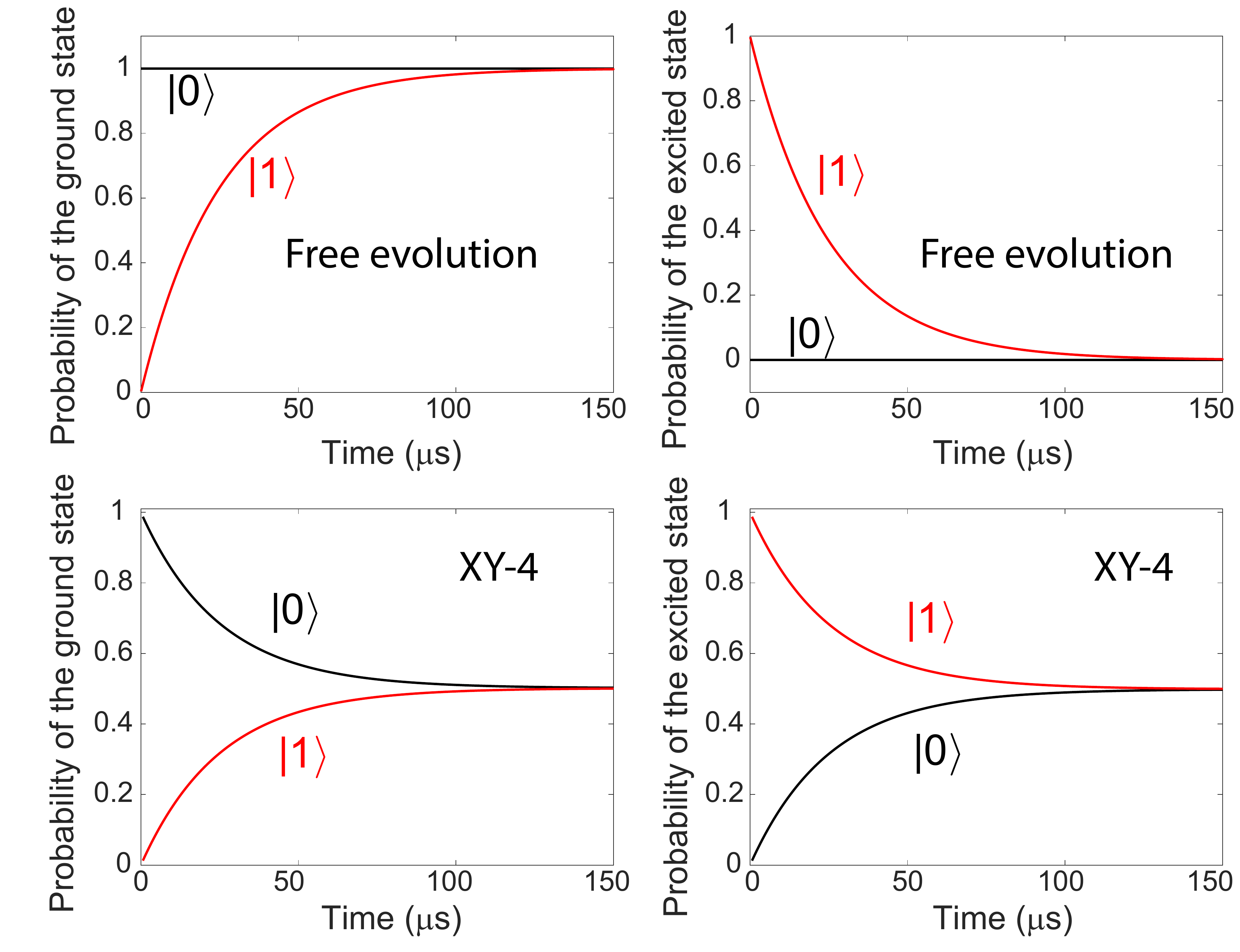}
\caption{Simulation of the quantum channel described by the equations \eqref{PDGAD}. The simulation starts in the ground state (black curves), $|0\rangle$, and the excited state (red curves), $|1\rangle$.  
\label{sim}}
\end{figure}

\begin{figure*}[t!]
\centering
\includegraphics[width=1 \linewidth]{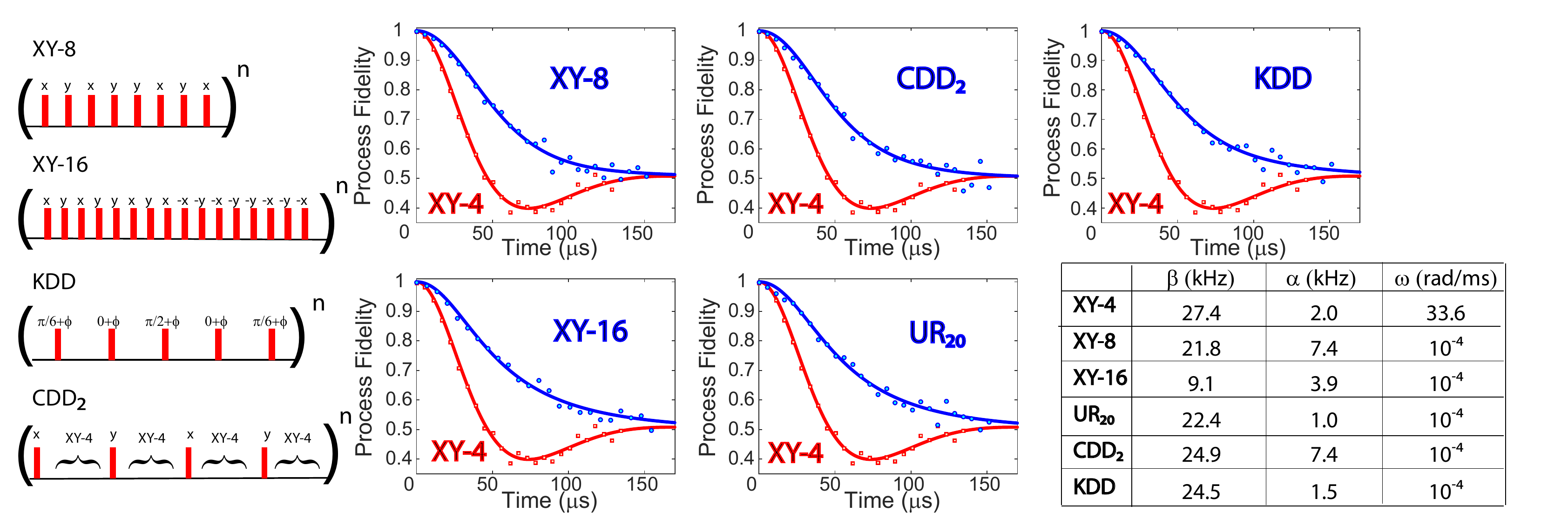}
\caption{Fidelity process as a function of time for different robust sequences, as indicated in each panel. The left panel illustrates the DD sequences tested. The delay between pulses is kept fixed at $\tau = 80 ns$ in all cases. The phase $\phi$ in the KDD sequence alternates between two values: $0$ and $\pi/2$. The basic cycle of the UR20 sequence, not shown, comprises 20 pulses with phases given by $\phi_k = (k-1)(k-2)\pi/10+(k-1)\pi/2$, where $k=1, \cdots,  20$ \cite{genov}. The table shows the fitting parameters for each sequence.   
\label{ddrobust}}
\end{figure*}

Figure \ref{echo} also shows that the observed fidelity decay for the Hahn echo is well fitted by the model \eqref{PDGAD}, which combines PD and AD channels. The fitting parameters are $\alpha = 20.9 kHz$ and $\beta = 23 kHz$. However, the $XY-4$ sequence cannot be fitted by the same model. The right panels of Figure \ref{echo} show the process matrices obtained at $t = 150 \mu s$ for the Hahn echo and the $XY-4$ sequence. Although the process matrix for the Hahn echo is fully compatible with the equations \eqref{PDGAD}, as $t \rightarrow \infty$, which maps any initial state to the equilibrium state $|0 \rangle$, the process matrix observed for $XY-4$ maps all initial states to a completely depolarized state. This result indicates that DD does not act to merely change coherence times.

We expect that after a long time interval, much greater than $T_1$, the qubit would return to the ground state. However, process tomography results showed that the final qubit state is a statistical mixture. This dynamics can be understood in terms of a model that results from combining DD and relaxation processes. When the qubit evolves freely, relaxation processes  drive it to the ground state. However, a DD pulse can remove the qubit from the ground state. Combining the continuous application of DD pulses with relaxation processes results in the statistical mixture of the ground and excited states, as the simulations in Figure \ref{sim} show. To model this dynamics, we consider a quantum channel that combines phase damping, bit-flip in different directions, and unitary rotations. A quantum channel that combines the above features is described by the Krauss operators 
\begin{eqnarray}
 K_1  &=&  \sqrt{\lambda(1-\gamma)} U, \\
 K_2 &=& \sqrt{(1-\lambda)(1-\gamma)} \sigma_z U , \\
 K_3 &=&  \sqrt{\frac{\gamma}{2}}   \sigma_x U , \\
 K_4 & =&  \sqrt{\frac{\gamma}{2}} \sigma_y U, 
  \end{eqnarray} 
where $ U = e^{-i\omega \Delta t \sigma_z/2} $, $1-\lambda = (1-e^{-\Delta t \alpha})/2$ is related to phase errors, as in \eqref{PDGAD}, and $\gamma = (1- e^{\Delta t \beta})$ is the probability of a bit-flip error occurs in the $x$ or $y$ direction. Figure \ref{echo}c shows that this model fits well the process fidelity for the $XY-4$ sequence. The fitting parameters are $\alpha = 2 kHz $, $\beta = 23 kHz $, and $\omega \approx. 10^{-2} rad/ms$ when the delay between pulses is $\tau = 120 ns$, $\alpha = 2 kHz $, $\beta = 27 kHz $, and $\omega = 33 rad/ms$ for $\tau = 80 ns$ and $\alpha = 7 kHz $, $\beta = 21 kHz $, and $\omega = 81 rad/ms$ for $\tau = 40 ns$.

Figure \ref{ddrobust} compares DD sequences. The $XY-4$ sequence has the worst performance among the sequences tested due to error accumulation. The oscillation of the fidelity observed for the $XY-4$ sequence can be associated with experimental imperfections \cite{souza6}, which cause the qubit to rotate around the $z$-axis. All robust sequences eliminate the oscillations. The $\alpha$ parameter relates to the rate that a phase error occurs; all DD sequences reduce this parameter roughly by a factor of ten when compared to the Hahn echo ($\alpha = 20.9 kHz$ ). This result clearly indicates that DD sequences are mitigating dephasing mechanisms. However, the $\beta$ parameter, which is related to relaxation processes that drives the qubit to the ground state, could not be reduced. Furthermore, combining DD and relaxation processes maps all states to a completely mixed state. All curves in Figure \ref{ddrobust} were fitted with the same effective channel described above.

\begin{figure}[h]
\centering
\includegraphics[width=1 \linewidth]{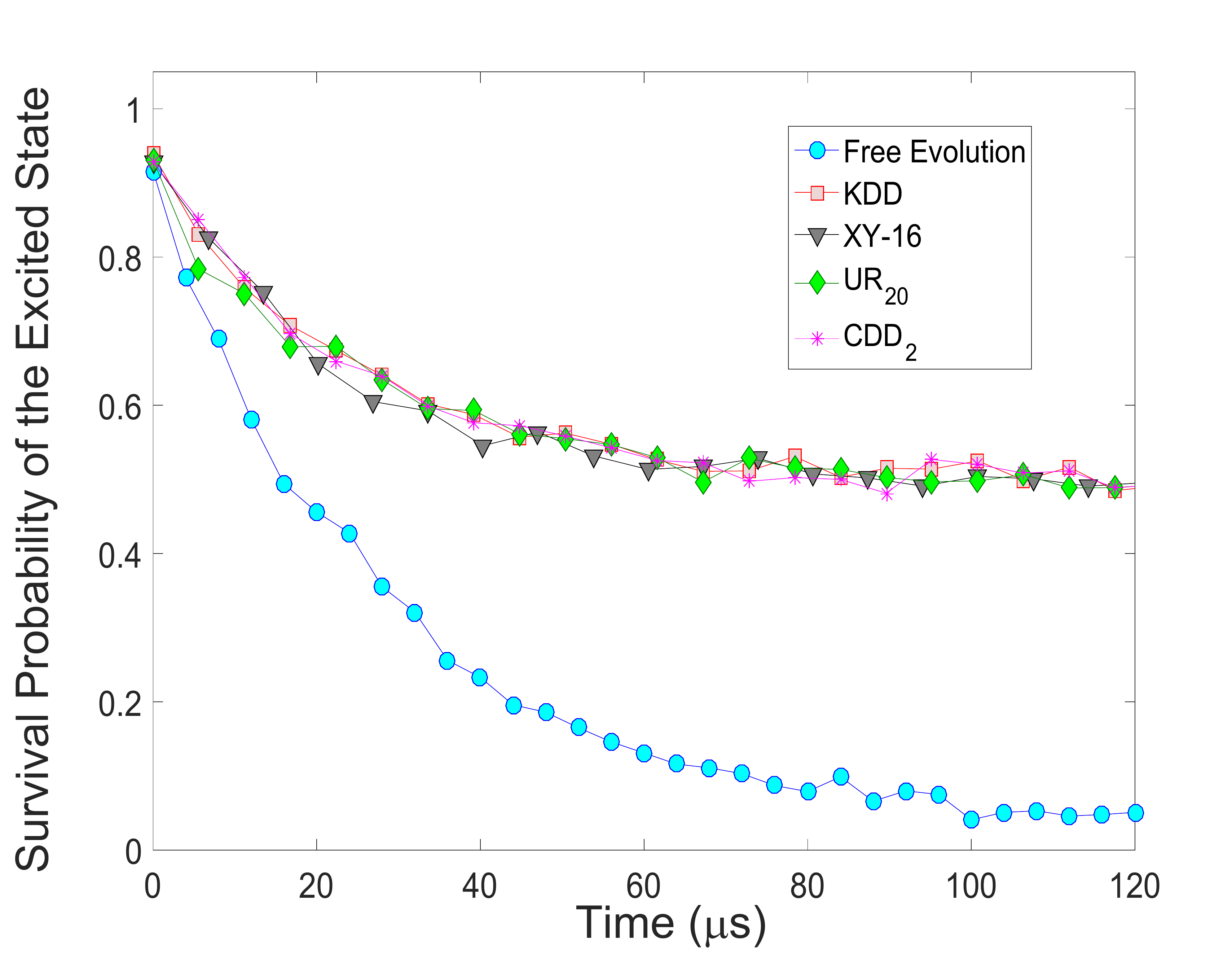}
\caption{Survival probability of the excited state $|1\rangle$ as a function of time.
\label{prop}}
\end{figure}

The survival probability of the excited state as a function of time is shown in Figure (\ref{prop}). At $t = 100 \mu s$, the survival probability is $\approx 0$, when we let the qubit evolve freely, and $\approx 0.5$, when DD is applied. This result appears to indicate that DD protects the state against spontaneous emission, as noted in \cite{lidar}.   However, DD appears to perform better here because it randomizes the qubit state,  the ratio in which the spontaneous emission occurs, as quantified by $\beta$, does not change significantly. Nonetheless, from Figures \ref{echo} and \ref{ddrobust}, we can conclude that using robust DD rather than non-robust sequences is preferable. 

\section{Conclusion} \label{conclusion} The ability to preserve quantum coherence in the presence of noise is essential for developing quantum devices. Dynamical decoupling is a widely used technique for protecting qubits against noise. Early experiments have demonstrated the usefulness of DD in many different types of qubits. Here, different DD sequences were tested in a cloud-based quantum computer. The performance of the sequences was characterized by quantum process tomography. Although decoherence and relaxation processes without DD are well modeled by a combined action of amplitude damping and phase damping channels, the application of DD results in different dynamics due to the combination of DD pulses and relaxation processes. As in early experimental tests of DD, the lowering of the sequence performance via the effect of pulse imperfections was shown here. Therefore, using robust DD sequences, designed to mitigate experimental errors, is preferred over non-robust sequences. The results reported here help in understanding the dynamics of superconducting qubits when DD is applied and demonstrate the usefulness of DD on the current available noisy quantum computers. An attractive prospect for future work is the implementation of DD in other platforms, such as the quantum computers based on trapped ions technology.

It must be noted that DD is effective only against low-frequency noise and does not by itself provide a route to scalable quantum computing. However, DD can be combined with error correction codes and quantum gates to improve the fidelity of the computation. The incorporation of DD sequences into quantum computing is not a trivial task. This must be done in such a way that the DD, which is designed to eliminate external perturbations, does not eliminate the control fields driving the computation. If the noise per gate is not high, it is possible to use DD to lower the resource requirements for quantum error correction, in this case, the advantages could be substantial \cite{west}. 

In the simplest case, we can make the computation insensitive to static perturbations by refocusing them with a Hahn echo \cite{shulman}. In the more general case, the Hahn echo has to be replaced by robust DD sequences. Initial experimental tests in this direction have been made recently on a Nitrogen-Vacancy Center \cite{sar,rong,souza1}, semiconductor quantum dot \cite{barthel}, and solid-state NMR \cite{souza2,souza3}. Possible schemes for incorporating  DD into quantum computing  were also suggested in \cite{viola,Khodjasteh1,Khodjasteh2,Khodjasteh3,west,Ng,capellaro,silva} and tested in liquid NMR \cite{boulant}. In most cases, these schemes were developed under the assumption that the DD sequence does not introduce any additional errors. Therefore, the implementation of such schemes with realistic control operations will require robust DD sequences.

\begin{acknowledgments}
This work was supported by the Brazilian National Institute of Science and Technology for Quantum Information (INCT-IQ) Grant No. 465469/2014-0, the National Council for Scientific and Technological Development (CNPq), and FAPERJ (Grant No. 203.166/2017). The author also acknowledges Rigetti computing for providing access to the quantum computing platform.
\end{acknowledgments}
 \bibliographystyle{apsrev}
  \bibliography{rigetti_dd}
\end{document}